\documentclass[aps,pra, amsmath, showpacs, preprintnumbers,superscriptaddress, twocolumn,sort&compress,floatfix, amssymb]{revtex4}
\pdfoutput=1
\usepackage{graphicx}
\usepackage{rotating}
\usepackage{dcolumn}
\usepackage{bm}
\usepackage{color}
\usepackage{mathptmx, textcomp}
\usepackage[latin1]{inputenc}
\usepackage{braket}

\usepackage{multirow}

\begin{document}

\author{M. Gr\"obner}\affiliation{Institut f\"ur Experimentalphysik und Zentrum f\"ur Quantenphysik, Universit\"at Innsbruck, 6020 Innsbruck, Austria}
\author{P. Weinmann}\affiliation{Institut f\"ur Experimentalphysik und Zentrum f\"ur Quantenphysik, Universit\"at Innsbruck, 6020 Innsbruck, Austria}
\author{F. Meinert}\affiliation{Institut f\"ur Experimentalphysik und Zentrum f\"ur Quantenphysik, Universit\"at Innsbruck, 6020 Innsbruck, Austria}
\author{K. Lauber}\affiliation{Institut f\"ur Experimentalphysik und Zentrum f\"ur Quantenphysik, Universit\"at Innsbruck, 6020 Innsbruck, Austria}
\author{E. Kirilov}\affiliation{Institut f\"ur Experimentalphysik und Zentrum f\"ur Quantenphysik, Universit\"at Innsbruck, 6020 Innsbruck, Austria}
\author{H.-C. N\"agerl}\affiliation{Institut f\"ur Experimentalphysik und Zentrum f\"ur Quantenphysik, Universit\"at Innsbruck, 6020 Innsbruck, Austria}

\title{A new quantum gas apparatus for ultracold mixtures of K and Cs and KCs ground-state molecules}

\date{\today}

\pacs{}

\begin{abstract}
 We present a new quantum gas apparatus for ultracold mixtures of K and Cs atoms and ultracold samples of KCs ground-state molecules. We demonstrate the apparatus' capabilities by producing Bose-Einstein condensates (BEC) of $^{39}$K and $^{133}$Cs in a manner that will eventually allow sequential condensation within one experimental cycle, precise sample overlap, and magnetic association of atoms into KCs molecules. The condensates are created independently without relying on sympathetic cooling. Our approach is universal and applicable to other species combinations when the two species show dramatically different behavior in terms of loss mechanisms and post laser cooling temperatures, i.e. species combinations that make parallel generation of quantum degenerate samples challenging. We give an outlook over the next experiments involving e.g. sample mixing, molecule formation, and transport into a science chamber for high-resolution spatial imaging of novel quantum-many body phases based on K-Cs.
\end{abstract}

\maketitle

\section{Introduction}
\label{Introduction}

Mixed quantum gases offer a wealth of research opportunities, ranging from precision measurements to studies of various quantum many-body regimes, some of which have no analogues in condensed matter physics. Questions on the fermionic effect on the mobility of bosons, the possibility of a supersolid phase for atomic Bose-Fermi mixtures \cite{bose_fermi_supersolid}, the existence of quantum phases that involve composite fermions \cite{Lewenstein_composite_fermions}, or mixtures under simultaneous superfluidity \cite{Salomon_bisuper} have attracted substantial theoretical and experimental interest. Bose-Fermi mixtures in bipartite lattices can facilitate understanding the interplay between interactions and topological band structures. Specifically Fermi-Fermi mixtures on a trimerized Kagome lattice open the possibility to study the physics of random valence bond frustrated quantum anti-ferromagnets, one of the paradigmatic problems of condensed matter physics \cite{quant_gas_kagome}.

Further, quantum simulation of exotic lattice models for the case of long-range interactions \cite{clustered_wigner,mols_trimerized_kagome}, with the added option to infer spatial correlations, ordering, and dynamics in such systems, seems to be on the horizon after the recent spectacular developments in the field of ultracold atoms. These include creation of ground-state lattice-trapped diatomic polar molecules \cite{Takekoshi2014,low_entropy_mols}, realization of optical lattices with non-cubic geometries \cite{Tarruell2012}, and high-resolution microscopy for both bosonic \cite{Greiner_microscope,Bloch_microscope} and fermionic \cite{KUHR_microscope,Zwierlein_microscope} atoms. The creation of an ultracold gas of molecules with low entropy is by itself challenging, but imperative given that the investigation of exotic quantum phases in most cases requires filling fractions near unity. Precise experimental control over the initial atomic constituents in a 3D lattice is necessary \cite{Danzl2010,low_entropy_mols}, followed by magneto-association and efficient coherent transfer to control the molecular ground state. High-resolution imaging techniques, although developed so far only for atoms, then invite similar realizations for polar molecules, either by using a predissociation step \cite{Danzl2008,Ni2008,Danzl2010}, or by applying direct imaging \cite{YeImaging} accompanied by cooling utilizing an appropriate semi-closed molecular transition \cite{DirectCoolInouye}. Additionally, manipulation of the dipolar long-range interaction by dressing the rotational molecular levels by DC or MW fields in 2D geometries provides a novel tool towards engineering of strongly correlated quantum phases \cite{Strongly_correlated_mols, Tailoring_molecules}. In a different context, MW/DC fields on a 2D lattice provide a toolbox for generating spin-1/2 XXZ models with direction-dependent spin anisotropy \cite{topological_ph_gorshkov,toolbox_Zoller,exoticmodels_Demler}, shown to exhibit symmetry-protected topological phases, and for implementing, on a honeycomb geometry, the Kitaev and Yao-Kivelson models \cite{kitaev_models}. One can also extend this control over to spin $S > 1/2$ interactions.

Last but not least, both the controlled adding and subsequent observation of the dynamics of impurities submerged in a Bose-Einstein condensate using a quantum gas microscope can provide another set of experimental handles on the physics of polarons inaccessible by conventional techniques \cite{Bloch_impurities}.

Alkali-metal atoms offer suitable combinations to approach the plethora of research directions laid out. Achieving quantum degeneracy in single-species alkali experiments is now a routine in many laboratories, however, achieving double degeneracy is not so common \cite{Modungo2002,Hadzibabic2002,Goldwin2004,Silber2005,Taglieber2008,Lercher2011,McCarron2011,Wu2011}. Evidently, the choice of combination is influenced also by the properties of the associated heteronuclear dimer molecule. A few combinations, namely NaK, NaRb, NaCs, KCs, and RbCs, are nominally stable  under two-body molecular collisions \cite{stable_chem}. The dimer $^{40}$KCs is the only choice of a chemically stable fermionic molecule apart from Na$^{40}$K. It is of interest due to its large dipole moment in its ground state of 1.9 D. Also, the two available potassium isotopes $^{39,41}$K increase flexibility and expand the possible scenarios for dimer association with Cs \cite{Patel2014}, in lieu of experiments employing bosonic molecules.

In this work we describe in detail a new K-Cs dual-species apparatus and the steps that are taken to obtain degenerate samples of $^{39}$K and Cs. In particular, we give details on the experimental timing sequence as this sequence has become comparatively complex. The apparatus is aimed at flexibly allowing K-Cs Bose-Bose and Fermi-Bose mixtures, the production of KCs bosonic or fermionic dipolar molecules, and imaging of such samples in a separate science chamber with single lattice-site resolution. The science chamber will be fed from a main collection chamber via optical transport of atoms in a moving focus dipole trap. We detail a conceivable scenario for producing ground-state KCs molecules. The substantial challenges in achieving individual condensation of $^{39}$K and Cs together with the large differences in post-laser cooling temperatures and evaporation paths make a parallel cooling-evaporation scheme difficult. We therefore adopt a sequential scheme with the capability to spatially displace the two samples in order to avoid two- and three-body losses involving $^{39}$K and Cs undergoing different incommensurate parts during their production cycle. In detail, we report on the first realization of Bose-Einstein condensation of K and Cs atoms in the same experimental apparatus. We obtain essentially pure, optically trapped BECs of Cs and $^{39}$K containing $4\times 10^4$ and $9\times 10^4$ atoms, respectively. We discuss the laser cooling strategy for both species and give a detailed description of the evaporation path that results in pure condensates. In particular we demonstrate condensation of $^{39}$K in a dipole trap at $\lambda = 880.25$ nm, which is the tune-out wavelength for Cs \cite{Arora11}, i.e. the wavelength at which Cs atoms in their ground state have vanishing AC polarizability. In a later stage of the experiment, this will give us a handle to overlap the K sample with the Cs sample without changes of the trap depth that might cause strong excitation for the Cs sample.

\begin{figure}
\includegraphics[width=1\columnwidth]{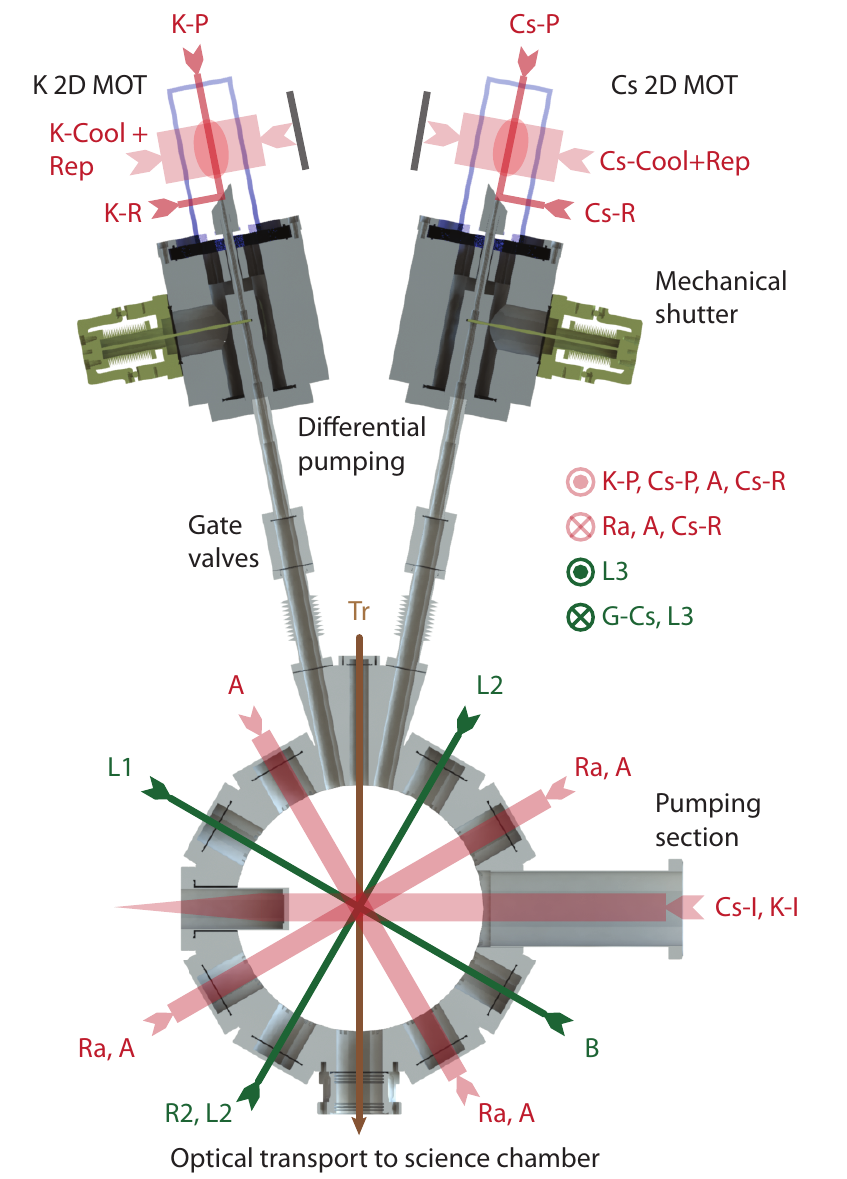}
\caption{\label{FIG1}Top view of the experimental setup. The two 2D$^{+}$ MOTs are at the top of the image, the main chamber is at the bottom. The following notation is used: Laser cooling beams for K: K-Cool+Rep cooling and repumping beams in 2D MOT, K-R retarding beam, K-P push beam; M-K-CR cooling and repumping beam for 3D MOT; D2-C, D1-C cooling and repumping beams for $D_2$, $D_1$ molasses cooling; K-P polarizer; K-I imaging; laser cooling beams for Cs: Cs-Cool+Rep cooling and repumping beams in 2D MOT, Cs-R retarding beam, Cs-P push beam; M-Cs-CR cooling and repumping beam for 3D MOT; Ra Raman beams; Cs-P polarizer; Cs-I imaging; Optical traps for Cs: R1, R2 reservoir; G-Cs guide; D-Cs dimple; Optical traps for K: D-K dimple, M-K 880.25-nm trap; Optical lattices L1, L2, L3; Tr optical transport beam; abbreviations A=(M-Cs-CR, M-K-CR, D2-C, D1-C) and B=(R1, L1, D-Cs, D-K, M-K).}
\end{figure}

\section{Experimental setup}
\label{Setup}

The experimental setup centers around a stainless steel main chamber in which laser and evaporative cooling towards condensation takes place (see Fig.~\ref{FIG1}). The outer diameter of the chamber is 28 cm. It has re-entrant viewports at the top and at the bottom (window diameter 16 cm) and 8 viewports with their normal direction in the horizontal plane for delivery of laser cooling, dipole trapping, and state-control light. Imaging is done through another re-entrant viewport via beams that are sent from the direction of the pumping section. The chamber is fitted with 4 parallel rod electrodes in rectangular configuration for the future generation of in-vacuum electric fields with amplitudes up to $20$ kV/cm (not shown in Fig.~\ref{FIG1}). It is fed by two independent 2D$^{+}$ magneto-optical traps (MOT) \cite{Dieckmann98}, one for each species. Note that previous experiments on BEC of Cs in our group in Innsbruck all have used the Zeeman slowing technique \cite{Weber2003,Rychtarik2004,Gustavsson2008} to provide an intense source of atoms. Here, we opted for the 2D-MOT technique that allows for a much more compact setup. The 2D$^{+}$ MOT chambers are connected to the main chamber by low conductivity tubes to achieve large differential vacuum levels ($\Delta p \textgreater 10^{-5}$ mbar). The 2D$^{+}$ MOTs are formed in indium-sealed rectangular quartz cells that operate at pressures around $10^{-6}$ to $ 10^{-7}$ mbar and act as a source of slow atoms. The pressures in these cells are dominated by the partial pressures of either K or Cs and are controlled by the source temperatures. The atomic beams cross in the middle of the main chamber. Here, the atoms are captured by standard 3D MOTs, serving as a first cooling stage. The atomic beams can be shuttered mechanically, but even with the two 2D$^{+}$ MOTs giving direct exposure to the main chamber, the pressure there is well below $10^{-10}$ mbar, as is confirmed by magnetic quadrupole trap lifetimes of over $40$ s. A comparatively complex set of magnetic coils generates the necessary B-fields and B-field gradients. Besides pairs of rectangular coils around the two 2D$^{+}$ MOT quartz cells, the setup features one pair of gradient coils, five independent pairs of offset coils that can easily deliver a combined field in excess of $1000$ G (e.g. two pairs of Bitter-type electromagnets \cite{Sabulsky13}, which allow for efficient water cooling), one Helmholtz pair of coils, and two Cosine coils. The Helmholtz and Cosine coils are needed to compensate for stray fields along the vertical and horizontal directions, respectively. All the laser light needed for dipole trapping, lattice generation, and optical transport is derived from one high-power $1064$-nm Nd:YAG single-frequency laser \footnote{Innolight Mephisto MOPA 42NE} with a maximum output power of $42$ W. During the different stages of the experiment the light of this laser is redirected into various beam directions as described below.

In the following, we provide data such as temperatures, beam waists etc. largely without specific errors. We estimate the typical statistical errors to be as follows: Temperatures, 2\%; beam waists, $1 \ \mu$m, laser beam powers, 10\%; particle numbers, 10\%, phase-space densities 20\%, magnetic field values, better than $10^{-4}$, and trap frequencies, 15\%. Systematic errors e.g. on the particles numbers might be significantly higher and will be subject to future detailed analysis.

\section{Formation of a cesium BEC}
\label{CsBEC}

\subsection{Cs laser cooling}
\label{CsLaser}

The initial collection and cooling of Cs atoms is achieved by conventional techniques \cite{Weber2003,Kraemer2004,Lercher2011}. Two independent home-built diode laser systems are locked to the $\ket{F=4}\rightarrow\ket{F'=5}$ and $\ket{F=3}\rightarrow\ket{F'=2}$ components of the $D_{2}$ transition, respectively, using modulation transfer spectroscopy on thermal Cs vapor in a glass cell. They provide the light needed for laser cooling of Cs in the 2D$^{+}$ MOT and 3D MOT and for repumping. Two tapered amplifiers (TA) give up to 160 mW of total cooling power for the 2D$^{+}$ MOT and 140 mW for the 3D MOT (as measured in front of the chambers after passing through acousto-optical modulators (AOM) for intensity and frequency control and optical fibers for beam delivery). The cooling light for the 2D$^{+}$ MOT is red detuned by about 12 MHz from the $\ket{F=4}\rightarrow\ket{F'=5}$ transition, while the significantly weaker repumping light is kept on resonance with the $\ket{F=3}\rightarrow\ket{F'=2}$ transition. A pure Cs sample from a broken ampule inside a vacuum nipple is heated to 30 °C and the resulting Cs vapor that diffuses into the quartz cell is cooled in two dimensions by two elliptically shaped retro-reflected beams in a $\sigma^{+}-\sigma^{-}$ configuration. To increase the flux and optimize the average speed for the atoms in the atomic beam, two laser beams (termed 'retarding' and 'push' beam) are overlapped along the longitudinal central axis of the 2D MOT. The retarding beam is reflected from an optical-quality polished stainless steel 45° wedge positioned inside the glass cell with a 1.5-mm-diameter hole in the middle for the atomic beam. The push beam, pointing towards the main chamber, creates a speed controlled atomic beam. By changing the power balance between these two beams we are able to optimize 3D MOT loading.

\begin{figure}
	\includegraphics[width=1\columnwidth]{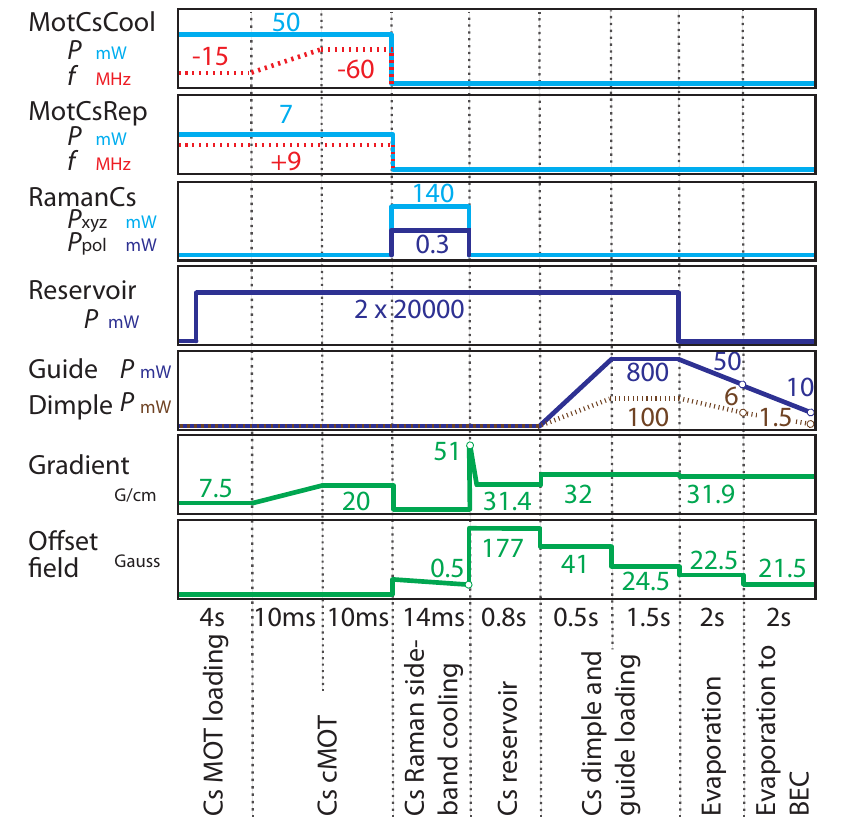}
	\caption{\label{FIG2} Typical timing diagram for Cs cooling. The continuous (blue, green) lines correspond to either total powers or magnetic field strengths and the dashed (red) lines to frequency detunings from the $\ket{F=4}\rightarrow\ket{F'=5}$ $D_{2}$ transition for the cooling and $\ket{F=3}\rightarrow\ket{F'=3}$ transition for the repumping light. Not shown is the timing for the 2D$^+$ MOT beams (Cooler: -12 MHz (50 mW along each axis, retro-reflected, intensity 4.5 mW/cm$^2$), Repumper: 0 MHz (6.5 mW, intensity 0.6 mW/cm$^2$)), which are simply switched off at the end of 3D MOT loading.}
\end{figure}

The loading rate into the 3D MOT in the main chamber is $2\times 10^8$ atoms/s. After $\sim 4$ s of loading time the 3D MOT has saturated at about $3\times 10^8$ atoms at a temperature of 115 $\mu$K. Next, the cloud is compressed by ramping up the magnetic field gradient from $7.5$ G/cm to $20$ G/cm within $10$ ms and holding this value for another $10$ ms (cMOT phase). Simultaneously, we change the detuning of the cooling light to $-60$ MHz while keeping the total power constant at 50 mW (the beam intensity in each beam is 2.7 mW/cm$^2$). To prepare the atoms for the next cooling step we optically pump them into the $\ket{F=3}$ state by switching off the repumper light a few ms before the cooling light. At this point, we have $2\times 10^8$ atoms at a temperature of 42 $\mu$K. To further laser cool the compressed cloud and to spin polarize the atoms in the Cs $\ket{F=3,m_{F}=3}$ absolute hyperfine ground state, we apply degenerate Raman-sideband cooling in an optical lattice \cite{Treutlein01,Weber2003} (we chose a 4 beam configuration for its inherent stability to vibrations). For the optical lattice we use light that before was used for the MOT and shift it into resonance with the $\ket{F=4}\rightarrow\ket{F'=4}$ transition. With a total power of 140 mW (250 mW/cm$^2$ for each beam) we create a three-dimensional optical lattice that drives Raman-sideband transitions and additionally acts as a repumper out of the $\ket{F=4}$ state. A polarizing beam (close to $\sigma^{+}$ polarization; power 0.3 mW (0.5 mW/cm$^2$); resonant with the $\ket{F=3}\rightarrow\ket{F'=2}$ transition) provides optical pumping as the atoms go down the lattice vibrational ladder to the targeted dark state $\ket{F=3,m_{F}=3}$. By applying an experimentally optimized field ramp for the offset field in the range of a few 100 mG we sweep through degeneracy the $\ket{n,m}$ and $\ket{n-1,m-1}$ manifolds ($n$ and $m$ are vibrational and Zeeman quantum numbers, respectively), remedying effects due to anharmonicity of the lattice potential and spatial variations of the lattice spacing. We polarize around 90\% of the atoms in the absolute ground state. After 14 ms of Raman cooling the temperature of the cloud with $5.5\times 10^7$ atoms is about 500 nK. The peak density and the phase-space density (PSD) are about $2.4\times 10^{11}$ cm$^{-3}$ and $2\times 10^{-3}$, respectively. These values offer perfect starting conditions for evaporative cooling in an optical dipole trap \cite{Weber2003}.

\subsection{Cesium dipole trap loading and evaporative cooling}
\label{CsDipole}

The laser-cooled Cs atoms are transferred to a large volume dipole trap (termed 'reservoir' trap \cite{Weber2003}) that is generated by two horizontally crossed 1064-nm beams (see Fig.~\ref{FIG1}) originating from a single beam in a bow-tie configuration. Care is taken to avoid interference effects by orthogonalizing the polarizations. The beams have a power of $20$ W, a waist of $550 \ \mu$m and are already switched on during the MOT loading phase (using mirror shutters actuated by bistable magnets). The reservoir trap has a depth of about $k_B \times 11 \ \mu$K. To load the Cs atom sample into the reservoir trap we ramp up the vertical magnetic field gradient and overlevitate the cloud for $2$ ms to counteract the effects of gravity after the Raman-sideband cooling. At the same time we also apply a 177-G offset field to increase the scattering rate (the Cs scattering length is now about $a_\mathrm{Cs}= 1850$ $a_{0}$ \cite{Weber2003}), and we estimate the collision rate to be $85$ 1/s. After the initial gradient kick we hold the gradient field at 31.4 G/cm and wait $800$ ms. The combined offset and gradient fields compensate for the effects of gravity but also result in a weak horizontal anti-trap. This anti-trap only plays a role for very weak traps at the end of the evaporation ramp and can be neglected at this stage \cite{Kraemer2004}. Although we use a large volume trap the temperature of the Cs sample increases to 1.7 $\mu$K. At this point we have about $1.5 \times 10^7$ atoms, fully polarized in the $\ket{F=3,m_{F}=3}$ state. Heating and atom loss can be largely attributed to imperfect phase-space matching. The Raman-sideband cooled cloud has a $1/e$-radius of around 385 $\mu$m, whereas an equivalent distribution of atoms in the reservoir trap would have a $1/e$-radius of 200 $\mu$m. Specific care has to be taken to avoid Landau-Zener transitions to other Zeeman sublevels during the transfer to the reservoir trap, which we do by adjusting our compensation coils. At last the deviation of the coils' geometry from the optimal Helmholtz configuration introduces trapping/anti-trapping in the horizontal/vertical planes, which also influences the efficiency of the transfer. Nonetheless, the inferred PSD stays almost constant.

The reservoir trap ensures high transfer efficiency from the laser cooling stage, but it is not suitable for efficient forced evaporative cooling, due to its low trap frequencies and the resulting low atomic densities. To overcome this problem, we use the 'dimple trick' and ramp up a narrow 43-$\mu$m dimple beam propagating along one of the horizontal axes to 100 mW and a 125-$\mu$m guide beam along the vertical axis to $800$ mW within $500$ ms, generating a tighter crossed trap. With this technique we locally change the shape of the trapping potential and increase the PSD of our sample \cite{Stamper98,Pinkse97,Weber2003}. During the ramp and a subsequent hold time of $1.5$ s the atoms in the large volume trap act as a reservoir and elastic collisions load atoms into the dimple without measurable heating of the reservoir sample. To maximize the atom number in the dimple trap and to control the three-body losses we reduce the scattering length during the ramp to $a_\mathrm{Cs}=900$ $a_0$ and during the hold time to $a_\mathrm{Cs}=380$ $a_0$. At the expense of a comparatively large number of atoms lost we locally increase the PSD to $8 \times 10^{-2}$. At this point, we still have $2.2 \times 10^6$ atoms in state $\ket{F=3,m_{F}=3}$ in the crossed trap and hence perfect starting conditions for forced evaporative cooling.

\begin{figure}
\includegraphics[width=1\columnwidth]{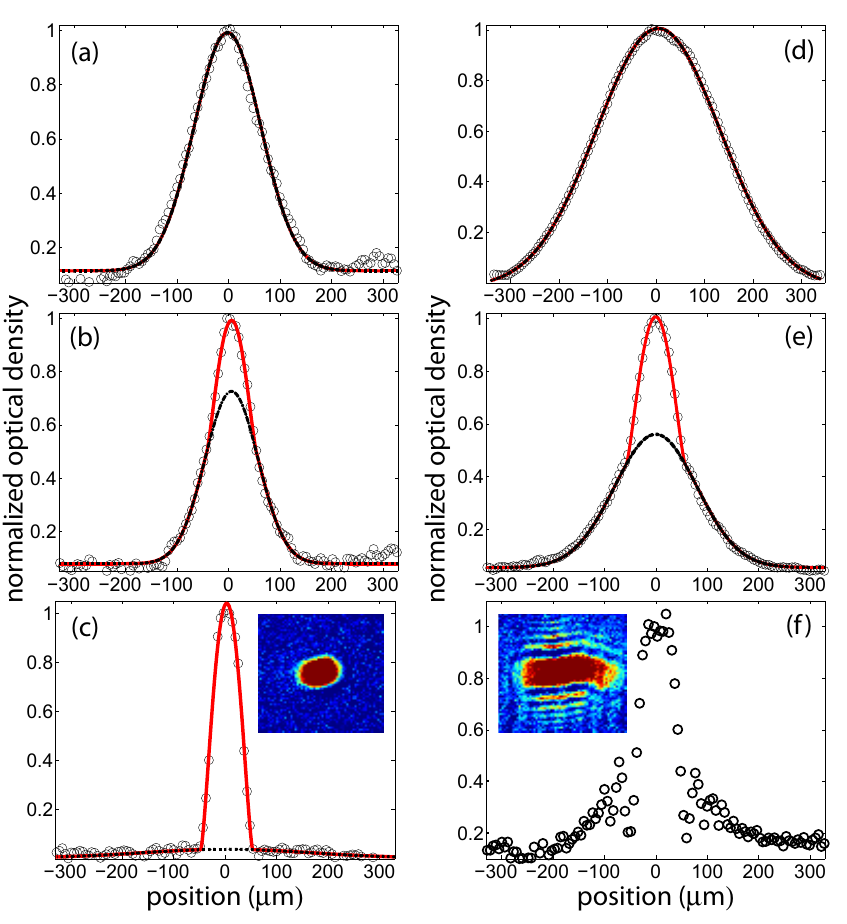}
\caption{\label{FIG3}Normalized optical density profiles showing the BEC phase transition for different stages of evaporative cooling for Cs (a-c) and $^{39}$K (d-f). The profiles are obtained by integrating the optical density along the horizontal axis. The solid (red) lines are fits using a bimodal distribution, while the dashed lines model the Gaussian-shaped thermal part. (a-c) are an average of four profiles taken after 60 ms of free expansion in the levitation field near zero scattering length. (d-f) are single profiles taken after 10 ms of free expansion at zero field. The insets in (c) and (f) show typical absorption images of essentially pure Cs and $^{39}$K BECs. For $^{39}$K one sees the distortion caused by the BEC convolved with lens aberrations.}
\end{figure}

We start forced evaporative cooling by switching off the reservoir trap within a few ms and by slowly ramping down the powers of the remaining guide and dimple beams. During the first 2-s-exponential ramp the powers of the guide and dimple beams are reduced to 50 and 6 mW, respectively. To suppress Cs three-body losses \cite{Weber2003b,Kraemer2006}, the magnetic offset field is set to $22.5$ G (giving $a_\mathrm{Cs}=290$ $a_0$) and the gradient is increased to 31.9 G/cm. We slightly over-levitate the Cs atoms to compensate for residual magnetic trapping along the vertical direction due to a deviation from the ideal Helmholtz configuration of the coils. After the first ramp, we still have a thermal cloud with a temperature of 55 nK. To achieve BEC we further ramp down the trapping beams to 10 and 1.5 mW and set the magnetic offset field to 21.5 G ($a_\mathrm{Cs}=245$ $a_0$). At the end of this 2-s ramp we observe an almost pure BEC of about $4 \times 10^{4}$ atoms as shown in Fig.~\ref{FIG3}(c).

The phase transition from a (thermal) Boltzmann gas to a BEC becomes evident in the appearance of a bimodal distribution, as clearly visible in time-of-flight absorption images and the corresponding vertical density profiles shown in Fig.~\ref{FIG3}(a-c). At higher trap depths the distribution is still thermal, showing a Gaussian shape as depicted in Fig.~\ref{FIG3}(a). Further reducing the trap depth, and respectively increasing the PSD, the central portion of the cloud assumes a parabolic (Thomas-Fermi) profile characteristic for a BEC (see Fig.~\ref{FIG3}(b)). By further applying evaporative cooling on the sample, the BEC fraction increases until the thermal component vanishes (see Fig.~\ref{FIG3}(c)). The density profiles are an average of 4 profiles measured after $60$ ms of free expansion in a levitating magnetic gradient field, while the scattering length is tuned close to zero to give the thermal component a chance to show up. For the result shown in Fig.~\ref{FIG3}(c) we calculate the in-trap Thomas-Fermi radii in radial and axial directions to be around 7 and 24 $\mu$m, respectively. The peak density of the condensate is $1.3\times 10^{13}$ cm$^{-3}$.

\section{Formation of a potassium BEC}
\label{KBEC}

\subsection{Potassium laser cooling}
\label{KLaser}

The initial 2D$^{+}$ MOT and 3D MOT stages to collect, trap, and cool $^{39}$K atoms are very similar to the ones described for Cs, but the non-resolved excited P$_{3/2}$ hyperfine structure of the $D_{2}$ transition demands a different laser cooling scheme to achieve sub-Doppler temperatures \cite{Gokhroo11,Landini11}. The core of the K laser system is a master laser locked to the $\ket{F=2}\rightarrow\ket{F'=3}$ $^{39}$K-$D_{2}$ cooling transition with an integrated TA that provides the cooling and repumping ($\ket{F=1}\rightarrow\ket{F'=2}$) light for the main laser cooling stages. The narrow $D_{2}$ excited-state hyperfine splitting demands the use of almost equal powers on the cooling and repumping transitions. After being diffracted by various AOMs, allowing for tunability and for switching between different isotopes, the light of the master laser gets amplified by four home-built TAs that provide $300$ mW of usable cooling power for repumping, for the 2D$^+$ MOT, and for the laser cooling in the main chamber. For the 2D$^+$ MOT the cooling and repumping light is red detuned by 29 and 15 MHz at experimentally optimized total powers of 140 mW (intensity 6.2 mW/cm$^2$ in each beam) and 50 mW (2.2 mW/cm$^2$ in each beam), respectively. To overcome low partial pressure at room temperature, we heat the K sample to 60 °C and the rest of the 2D$^{+}$ MOT chamber to 40 °C.

\begin{figure}
\includegraphics[width=1\columnwidth]{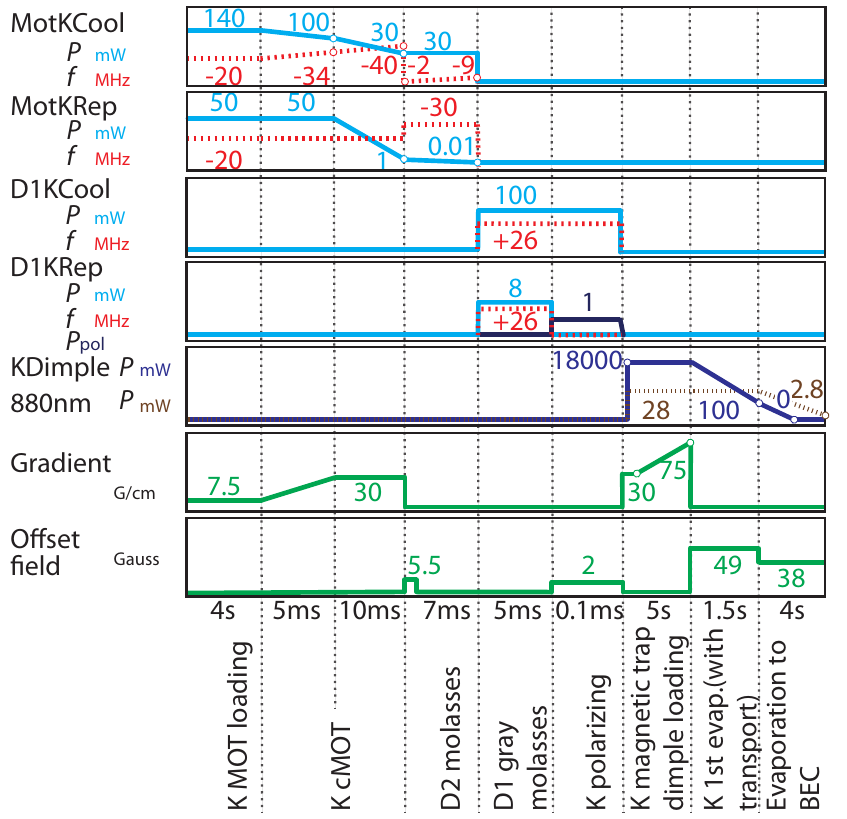}
\caption{\label{FIG4}Typical timing diagram for the $^{39}$K cooling sequence. The continuous (blue, green) lines correspond to either total powers or magnetic field strengths and the dashed (red) lines to frequency detunings from the $\ket{F=2}\rightarrow\ket{F'=3}$ $D_{2}$ transition for the cooling and $\ket{F=1}\rightarrow\ket{F'=2}$ transition for the repumping light. Not shown is the timing for the 2D$^{+}$ MOT beams (Cooler: -12 MHz (70 mW along each axis, retro-reflected, intensity 6.2 mW/cm$^2$), Repumper: 0 MHz (25 mW along each axis, intensity 2.2 mW/cm$^2$)). These are switched off at the end of the 3D MOT loading phase.}
\end{figure}

As in the Cs case we first capture the $^{39}$K atomic beam in a 3D MOT, which saturates after 4 s of loading at $8.5\times 10^8$ atoms. The subsequent timing is shown in Fig.~\ref{FIG4}. We find that optimum collection happens at a magnetic field gradient of 7.5 G/cm and for cooling beams (23 mW in each of the 6 beams; intensity 7.7 mW/cm$^2$) and repumping beams (8.3 mW; 3 mW/cm$^2$) that are both red-detuned by $20$ MHz. By increasing the gradient to 30 G/cm, reducing the total powers of the cooling and repumping beams to 30 and 1 mW, and increasing the red detuning from the cooling transition to 40 MHz, we compress the cloud (cMOT) while suppressing light-assisted collisions that otherwise heat the atomic sample. At this point, we detect $8\times 10^8$ atoms at a temperature of 2.2 mK. To further laser cool the $^{39}$K atoms, sequential optical molasses cooling, first on the $D_2$, then on the $D_1$ transition is applied. In general, optical molasses cooling works very efficiently when the excited-state hyperfine splitting is much larger than the natural linewidth and one can far red-detune the light from the cycling transition. For $^{39}$K this is not possible and the cooling strategy on the $D_2$ line is very different \cite{Gokhroo11,Landini11}. After the compression the cooling-light frequency is slightly detuned to the red of the $\ket{F=2}\rightarrow\ket{F'=3}$ transition and ramped from -2 to -9 MHz in the course of the cooling stage. During the ramp it is sufficiently far blue detuned from the $\ket{F=2}\rightarrow\ket{F'=2}$ transition to realize sub-Doppler cooling and to avoid negative friction forces. At the same time, the total repumper power is ramped down to very small values (from 1 to 0.01 mW) and its frequency is detuned far to the red (-30 MHz). An offset field pulse (1 ms, 5.5 G) is applied in the very beginning of the $D_2$ molasses cooling stage in order to compensate for stray magnetic fields occurring from switching off the gradient field. With this technique we reach temperatures down to 30 $\mu$K.

To achieve even lower temperatures and to polarize the atoms into the magnetically trappable $\ket{F=1,m_{F}=-1}$ state, we take advantage of the $^{39}$K $D_1$ line and its well-resolved excited hyperfine states and use a gray molasses cooling scheme \cite{Nath13,Salomon13}. The light is derived from a home-built master laser that is directly locked to the crossover transition of the $D_1$ line using modulation transfer spectroscopy. The laser light is then modulated by a resonant electro-optical modulator (EOM), which imprints sidebands at the ground-state splitting of $461.7$ MHz. The modulated light is amplified by a home-built TA. To induce gray molasses cooling we tune the repumper and cooling lasers to the Raman resonance in $\Lambda$-configuration and detune both of them to the blue of the P$_{1/2}$ $\ket{F=2}$ state (+26 MHz). With total powers of 100 mW (intensity 5.3 mW/cm$^2$ in each beam), we can decrease the temperature of the sample to 7 $\mu$K in 5 ms without observing any significant atom loss. To polarize the sample, we apply a 2-G offset field and a $\sigma^-$ polarized beam on the $\ket{F=1}\rightarrow\ket{F'=1}$ $D_1$ transition (46 $\mu$W, 0.2 mW/cm$^2$), while keeping the cooling light still on. The atomic population is then shelved into the $\ket{F=1,m_{F}=-1}$ dark state. This gives us a factor of 2 more atoms in state $\ket{F=1,m_{F}=-1}$ at the expense of a slight temperature increase that depends on the density of the cloud. We optimize the polarization procedure parameters mainly on the final number of atoms loaded into the subsequent dipole trap.

\subsection{Magnetic trapping and evaporative cooling in a dipole trap}
\label{KDipole}

Next, we implement a magnetic quadrupole trap, with an initial gradient of 30 G/cm, to filter the atoms not residing in the $\ket{F=1,m_{F}=-1}$ state due to insufficient optical pumping. Starting from $7\times 10^8$ atoms after $D_1$ molasses cooling, we load $5\times 10^8$ atoms with a temperature of 43 $\mu$K into the magnetic trap. Given the background scattering length of $a_K=-33$ $a_0$ for $^{39}$K, which gives rise to a Ramsauer-Townsend minimum in the collisional cross section at 400 $\mu$K \cite{Landini11}, direct evaporation in the magnetic trap becomes very inefficient. To overcome this limitation, we overlap a two-color (880.25 nm, 1064 nm) dipole trap beam with the $^{39}$K sample during the whole magnetic-trap time. To increase the elastic collision rate in our system we are bound to increase the collision energy to values above the Ramsauer-Townsend minimum. We use a 20 $\mu$m waist with a total power of 18 W at 1064 nm (the 880.25-nm beam with 28 mW becomes important only in the end of the evaporation stage). This tightly focused beam increases the collisional rate inside the optical trap and allows for efficient loading. To further increase the atom number in the dipole trap we also linearly compress the magnetic trap to a maximum gradient of 75 G/cm.

After 5 s of combined magnetic trapping and dipole trap loading, we switch off the magnetic trap and exponentially ramp down the power in the 1064-nm trap to 100 mW within 1.5 s. To allow forced evaporative cooling and to tune the $^{39}$K scattering length, we use the region between the two intraspecies Feshbach resonances at $33.6$ and $162.3$ G \cite{Errico07}. During the first ramp we choose a scattering length of $a_{\mathrm{K}} = 57$ $a_0$ to suppress three-body losses. In the second ramp we linearly ramp down the 1064-nm dipole trap to zero power in 2 s, while also exponentially ramping down the 880.25-nm trap to $2.8$ mW in 4 s. The lower intensities and hence reduced densities during this ramp allow us to choose a higher value for the $^{39}$K scattering length and we find the optimum value at $a_{\mathrm{K}}=225$ $a_0$ at $B=38$ G. Fig.~\ref{FIG3}(d-f) shows the phase transition from a thermal cloud to what we believe to be an essentially pure BEC in state $\ket{F=1,m_{F}=-1}$. We observe the density profiles after $10$ ms of free expansion at zero field. Fig.~\ref{FIG3}(d) shows the Gaussian-shaped thermal density distribution at a temperature of $980$ nK measured after the first evaporation ramp. By further decreasing the trap depth we cross the BEC transition with around $1\times 10^6$ atoms and obtain images as the one shown in Fig.~\ref{FIG3}(e). The integrated profile is well fit by a bimodal distribution. At this point, the 1064-nm dipole trap beam still has a power of 50 mW. From the thermal component we obtain a temperature of $214$ nK. Further deep evaporation (extinguishing the 1064-nm light, ramping the 880.25-nm light to $2.8$ mW) leads to images as the one shown in the inset to Fig.~\ref{FIG3}(f). We observe strong fringing that we attribute to a lensing effect due to the presence of a dense condensed sample in combination with aberrations caused by the imaging system \cite{Bradley95,Bradley97}. The final trapping frequencies of our anisotropic single-beam trap are $\omega_{z}/2\pi=3$ Hz and $\omega_{r}/2\pi=254$ Hz. In this regime the chemical potential $\mu$ is on the order of $\hbar\omega_r$ and we are thus entering the 1D-3D crossover regime in which the Thomas-Fermi approximation is not directly applicable. By following the approach described in Ref. \cite{Gerbier04} we calculate the radius of the condensate to be $R=1.8$ $\mu$m and the half-length to be $L=249$ $\mu$m. The tight confinement in the radial dimension makes a direct analysis of the absorption images difficult. In future experiments we will aim at more spherical K-BECs to allow for better spatial overlap with the Cs sample. We estimate the number of particles to be $9\times 10^4$ in what we think is an essentially pure BEC.

\section{Conclusions and Outlook}
\label{DualBEC}

\begin{figure}
\includegraphics[width=1\columnwidth]{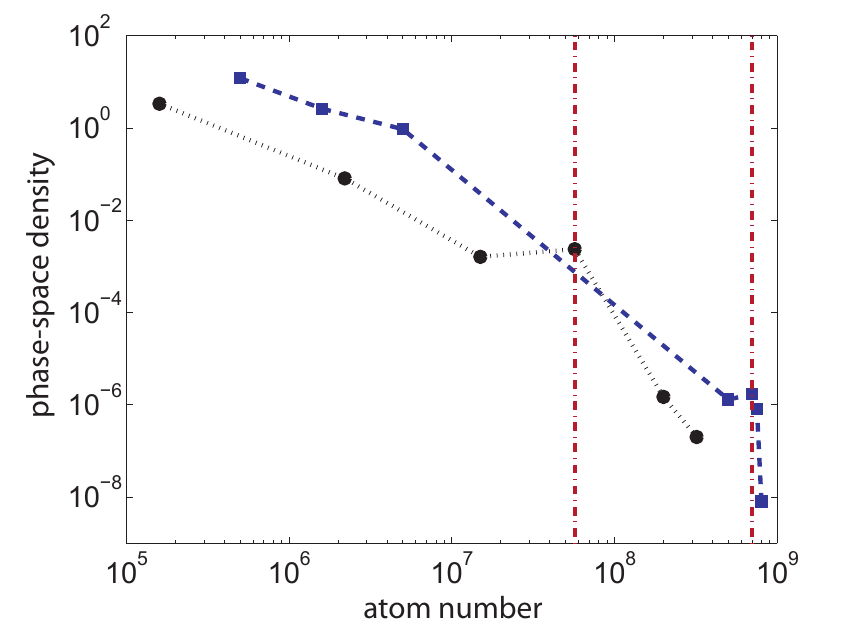}
\caption{\label{FIG5} Measured phase-space density assuming a Boltzmann gas during laser and evaporative cooling as a function of the atom number. The filled black circles represent typical PSDs during a Cs cooling sequence. From right to left: 3D MOT, cMOT, Raman-sideband cooling, reservoir trap, dimple and guide at 100 and 800 mW, first evaporation ramp; The filled (blue) squares show the same for $^{39}$K. From right to left: cMOT, $D_{2}$ molasses, $D_{1}$ molasses, magnetic trap, dipole trap at 200, 100, and 50 mW; The vertical dash-dotted (red) lines indicate the end of laser cooling for Cs (left) and for $^{39}$K (right).}
\end{figure}

In this paper we have demonstrated the independent realization of BECs for Cs and $^{39}$K in one experimental apparatus. We efficiently condense both species without any additional coolant and realize essentially pure condensates with $4\times 10^4$ and $9\times 10^4$ atoms, respectively. The path through phase space as a function of particle number, as shown in Fig.~\ref{FIG5}, illustrates the different challenges one faces working with $^{39}$K and Cs. For Cs, laser cooling works very efficiently and results in a spin-polarized sample with a PSD of $2.4\times 10^{-3}$. The challenge here is the reservoir and crossed dimple/guide dipole trap loading. One has to choose carefully the dimensions of the dipole traps involved. Small waists (e.g. a 20 $\mu$m dimple) lead to too high densities and enhanced three-body losses, while large ones prohibit efficient evaporative cooling. For $^{39}$K the situation is very different. The narrow hyperfine structure in the $D_2$ excited state prevents efficient sub-Doppler cooling and makes the use of an additional $D_1$-line cooling stage preferable. The two-stage molasses cooling results in a non-polarized sample with a comparably low PSD of $1.8\times 10^{-6}$. Consequently, we have to polarize and filter the sample by using a magnetic trap and to load a very high intensity dipole trap to overcome the low-lying Ramsauer-Townsend minimum.

The challenge to work with the two species simultaneously becomes even more apparent when one compares their individual Feshbach spectra. Both species offer a variety of different resonances in the low- and high-field region \cite{Berninger13,Errico07} that allow for a precise control of their scattering lengths. The BECs presented in this paper are realized in the low-field region, where efficient simultaneous evaporation of both species seems to be difficult because condensation is achieved in non-overlapping magnetic field regions only (near $21.5$ G for Cs and near $38$ G for $^{39}$K). One way is to take advantage of the fact that overlapping broad Feshbach resonances exist for $^{39}$K in the $\ket{F=1,m_{F}=-1}$ state and for Cs with poles at 560.7 and 548.8 G, respectively. Close to these two resonances one should be able to limit the three-body losses for Cs and to tune the scattering length for $^{39}$K to around 200 $a_0$. We are currently preparing high-field absorption imaging for diagnosis in this magnetic field region. Alternatively, we are installing an optical lattice that will allow us to 'park' one of the two species in a one-atom Mott insulator \cite{Mark2011,Meinert2013} while the other species is cooled towards degeneracy. In the one-atom Mott insulator the atoms are protected from colliding with each other. Storage in such a state should be possible for many seconds \cite{Mark2012}.

Generally, the sequences presented for $^{39}$K and Cs are designed to allow for spatial overlap of the two atomic species. Especially the trapping approach for $^{39}$K using a double frequency dipole trap, where we have full control over the 3D position of the K sample (AOM in single-pass configuration for horizontal axis control, piezo mirror for vertical axis control, translation stage for control along the longitudinal axis), gives us an efficient handle to overlap both samples. One possibility is to start with the $^{39}$K cooling sequence as before (Fig.~\ref{FIG4}), but then to interrupt forced evaporation after just one ramp when the atoms in the dimple trap are at $T\sim 2 \ \mu $K. During that first phase of evaporation we longitudinally displace the cloud from the MOT center to avoid two- and three-body losses with the Cs atoms, which now will be starting their MOT phase and which will be highly energetic. Stored in this sharply focused trap, the $^{39}$K sample is very robust against changes in magnetic offset and gradient fields, which allows us to start with the Cs sequence identical to the one discussed previously (Fig.~\ref{FIG2}). When this sequence reaches the reservoir phase the offset fields are such that we can efficiently proceed with the second evaporation ramp for $^{39}$K. For the third ramp of $^{39}$K, which now coincides with the first for Cs, one has to choose a magnetic field value that serves as a compromise and that works for both species. We expect to be able to identify a window in the vicinity of 550 G to achieve simultaneous degeneracy. At the end or during that last evaporation stage the $^{39}$K cloud is brought back spatially, now residing only in the 880.25-nm trap, and it can be overlapped with the Cs cloud. The overall sequential procedure has been tested and gave encouraging results with the missing link that for the last ramps we chose an offset field to condense either $^{39}$K or Cs. Again, parking one of the species, e.g. Cs, in a one-atom Mott insulator is also an option.

The strategy described here allows us in a next step to explore the experimentally not confirmed interspecies scattering properties of K and Cs \cite{Patel2014}. Initially, we will perform Feshbach spectroscopy, searching for Feshbach resonances as evidenced by increased three-body loss (for similar experiments performed in our group on ultracold Rb-Cs mixtures and samples of ultracold RbCs molecules, see Ref.'s~\cite{Pilch2009,Takekoshi2012}). Based on such measurements we will attempt KCs Feshbach association, first in free space \cite{Herbig2003,Mark2005}, and later with more control when Cs is already loaded into an optical lattice at unity filling. A similar strategy is currently pursued within our group's RbCs project. Details are laid out in Ref.~\cite{Lercher2011}. To subsequently produce ultracold samples of ground-state molecules \cite{Danzl2008,Ni2008,Danzl2010} we plan on using stimulated Raman adiabatic passage (STIRAP) using a single $\Lambda$-configuration with branches at approximately 1000 nm and 1500 nm as suggested in \cite{Stirap_KCs} and as has already been mastered in our laboratory using RbCs \cite{Takekoshi2014}. We have set up two filter-cavity diode-laser systems capable of achieving linewidths in the vicinity of $1$ Hz when locked to a vacuum-enclosed stable ultra-low expansion (ULE) reference resonator \cite{cavity_laser} with the aim to reach STIRAP efficiencies of 90\% and above.

In a further experimental line, we will optically transport the atomic samples into a science chamber (to be attached to a gate valve on the main chamber opposite to the 2D$^+$ MOTs, see Fig.~\ref{FIG1}), where we are ready to install a high-resolution home-built microscope, achromatic for the $D_{2}$ wavelengths of both species with a numerical aperture of NA=0.64. The science chamber, a small stainless-steel apparatus ready for operation, is also equipped with electrodes for applying electric fields to polarize the molecules. It is intended for single-site resolved imaging of atomic Cs and K, K-Cs mixtures, and KCs molecules.

Next to the bosonic research line discussed here, our experimental setup is ready to work with the fermionic $^{40}$K isotope. The K sample in the 2D$^+$ MOT source is enriched to 9\% with $^{40}$K. So far we have only tested the source sample and the laser system by realizing a $^{40}$K 2D$^+$ MOT. For the near future we will benefit from our experience with $^{39}$K and implement all the relevant laser cooling steps on the fermionic isotope, e.g. $D_1$ molasses cooling \cite{Fernandes12}.

We are indebted to R. Grimm for generous support. We thank M. J. Mark for contributions at an early stage of the project, M. Landini for discussions, and acknowledge contributions to the 2D$^{+}$ MOT setup by B. Ziernh\"old, to the STIRAP laser setup by M. Segl, and to the high-resolution imaging setup at the science chamber by M. Marszalek. We gratefully acknowledge funding by the European Research Council (ERC) under Project No. 278417 and by the Austrian Science Foundation (FWF) under Project No. I1789-N20 (joint Austrian-French FWF-ANR project).

\end{document}